# The digital labour of artificial intelligence in Latin America: a comparison of Argentina, Brazil, and Venezuela

Paola Tubaro[1], Antonio A. Casilli[2], Mariana Fernández Massi[3], Julieta Longo[4], Juana Torres Cierpe[5], Matheus Viana Braz[6]

(1) Corresponding author. Centre of Research in Economics and Statistics (CREST, CNRS-ENSAE), Polytechnic Institute of Paris, France, paola.tubaro@cnrs.fr. ORCID: 0000-0002-1215-9145

(2) DiPLab research programme, Telecom Paris | Polytechnic Institute of Paris, France, antonio.casilli@ip-paris.fr. ORCID: 0000-0003-2025-1627

(3) Instituto de Investigaciones en Humanidades y Ciencias Sociales (IDIHCS-CONICET), Argentina, marianafmassi@gmail.com. ORCID: 0000-0002-7379-1507

(4) Instituto de Investigaciones en Humanidades y Ciencias Sociales (IDIHCS-CONICET), Argentina, longojulieta@gmail.com. ORCID: 0000-0002-7896-8965

(5) Inria, France, juana.torres-cierpe@inria.fr. ORCID: 0009-0005-3426-5167

(6) Department of Psychology, Brazil Minas Gerais State University (UEMG), Brazil, matheus.braz@uemg.br. ORCID: 0000-0003-1193-9753

## Abstract

The current hype around artificial intelligence (AI) conceals the substantial human intervention underlying its development. This article lifts the veil on the precarious and low-paid 'data workers' who prepare data to train, test, check, and otherwise support models in the shadow of globalized AI production. We use original questionnaire and interview data collected from 220 workers in Argentina (2021-22), 477 in Brazil (2023), and 214 in Venezuela (2021-22). We compare them to detect common patterns and reveal the specificities of data work in Latin America, while disclosing its role in AI production.

We show that data work is intertwined with economic hardship, inequalities, and informality. Despite workers' high educational attainment, disadvantage is widespread, though with cross-country disparities. By acknowledging the interconnections between AI development, data work, and globalized production, we provide insights for the regulation of AI and the future of work, aiming to achieve positive outcomes for all stakeholders.

## Funding details

This work was supported by Mission pour les Initiatives Transverses et Interdisciplinaires of the French National Centre for Scientific Research (MITI-CNRS) as part of its *Enjeux de l'IA* challenge (2020-21), by Maison des Sciences de l'Homme Paris Saclay under grant 20-MA-02, and by Minas Gerais Research Fondation, FAPEMIG (FAPEMIG/UEMG program, 05/2021).

## Disclosure statement

In compliance with the General Data Protection Regulation (GDPR), the data collection and treatment were submitted and approved through CNRS's Data Protection Officer. All participants provided written informed consent.

The authors report there are no competing interests to declare.

## Authors' biographical notes

**Paola Tubaro** is research professor in sociology and technology at the National Centre for Scientific Research in Paris. She has extensively published on the platform economy, the social and ethical dimensions of artificial intelligence, social and organizational networks, and data methodologies. She is leading several projects on inequalities and pauperization among online micro-workers.

**Antonio A. Casilli** is professor of sociology at Polytechnic Institute of Paris. His research foci are digital labour, data governance, and human rights. He co-founded the research program DiPLab (Digital Platform Labor) and the INDL (International Network on Digital Labor). Among his publications, the award-winning book *Waiting for Robots* (University of Chicago Press, 2025).

**Mariana Fernández Massi** is a tenure researcher at the National Scientific and Technical Research Council (CONICET) and lecturer at the Universidad Nacional de Moreno in Argentina. Her research interests focus on changes in working time and space linked to automation and digitalisation, and trade union responses to these transformations.

**Julieta Longo** is a tenure researcher at the National Scientific and Technical Research Council (CONICET) and lecturer at the Universidad Nacional de La Plata in Argentina. Her areas of research concern working conditions and the organisation of working time and space for freelance workers.

**Juana Torres Cierpe** is a labour sociologist at the French National Institute for Research in Digital Science and Technology (INRIA). She specializes in topics related to platform labour in Latin America and work behind artificial intelligence. Currently, she is analysing the challenges of AI implementations among workers in the French public administration.

**Matheus Viana Braz** is psychologist and professor in the Department of Psychology at Maringá State University (UEM), Brazil. His current research is centred on the platform economy and digital inequalities within artificial intelligence production, with particular emphasis on Latin America.

# Introduction

The recent emergence of artificial intelligence (AI) as a transformative force across science and industry conceals the substantial human intervention underlying its development. Beyond high-earning Silicon Valley engineers, the human inputs to the production of speech recognition, text generation, computer vision, and other state-of-the art technologies include masses of 'data workers' who perform unglamorous supporting tasks. They prepare data to train machine-learning algorithms, for example by highlighting features (eyes, nose, mouth) and emotions (happiness, anger) in photographs to be used in the development of face recognition technologies. Data workers also check the outputs of AI tools to monitor their correct functioning, and even replace algorithms when tasks are hard to automate, for example to detect contents in a blurred image (Tubaro et al., 2020).

Fragmented and repetitive, these tasks are deemed peripheral to the core value-adding activities of AI producers and are commonly outsourced via online labour platforms and/or networks of subcontractors. The contribution of data work remains largely invisible and attracts low remunerations, often by piecework and without any long-term commitment. Ironically, its very existence challenges gloomy narratives of job destructions through AI, but it comes with deteriorated working conditions and low pay. Caught in its internal contradictions, data work appears as an instance of 'digital labour', a technology-fuelled reconfiguration of human productive activity that induces a progressive shift of the balance of power away from labour and toward capital. Ekbia and Nardi (2017) call 'heteromation' the capital accumulation based on extraction of economic value from low-cost labour in computer-mediated networks. Its enablers are technical infrastructures that handle worker management as if it were a computational problem, thereby disguising the employment nature of the relationship and the related human subjectivities and rights. Organizational arrangements compound the effects of technical framings insofar as outsourcing excludes data workers from the resources of lead technology companies and weakens the capacity of regulative institutions to offer protection against market fluctuations (Wood et al., 2019).

In scholarly and policy debates, data work is often lumped in with 'uberised' digital labour as often seen in ride-hailing and delivery services. Like these other emerging forms, it largely escapes the rules of salaried employment and raises concerns about erosion of job security and social protection (Berg et al., 2018). But apparent commonalities hide deeper differences: unlike ride-hailing and delivery, data work can be performed remotely with a computing device and internet connection. AI producers, mostly located in richer countries, therefore have strong incentives to offshore data tasks to cheaper contributors overseas (ILO, 2021). If data work was first documented in the United States in the late 2000s (Ross et al., 2010), it is now traded in a planetary market encompassing at least 75 countries (Berg et al., 2018), and the majority of labour supply now originates in lower-income regions (ILO, 2021). As an input to AI development, data work must be understood in light of the cross-country dependencies, market asymmetries, and power dynamics that in recent decades, have driven the unbundling of much industrial production through worldwide trade (Milberg & Winkler, 2013). A narrow focus on challenges to (national) labour law would miss these broader effects.

Taking data work as a pivotal factor in the global political economy of AI, this article aims to uncover the social, political, and economic forces that shape it in under-researched geographies, notably in Latin America where it is massively present according to recent, albeit unsystematic evidence (for example, Grohmann & Araújo, 2021; Miceli et al., 2020; Posada, 2022). Who are the data workers in the region, and how do they join this market? How does participation differ (if at all) across countries, and with what implications on employment, growth, and inequalities? To what extent can data work trigger development, moving countries up the global AI supply chain?

To address these questions, we lift the veil on the people who perform data tasks in the shadow of globalized AI production in three major Latin American countries: Argentina, Brazil, and Venezuela. We use original mixed-method data (questionnaires and in-depth interviews) to compare and contrast these cases in order to reveal common patterns and expose the specificities that distinguish Latin America. We also disclose the precise role of data work in feeding contemporary AI production and derive policy implications at global and national levels.

## Data work for AI: a planetary trade

Data work can be as trivial as recording one's voice while reading aloud a sentence, and as complex as assigning each pixel in a picture to a given object (for example, hair, lips, etc. in images of faces). It supports AI through the three functions of 'preparation' (data generation and annotation), 'verification' (performing manual output checks at the end) and 'impersonation' (replacing algorithms when they fail) (Tubaro et al., 2020). A landmark example is the ImageNet database, which was instrumental in revealing the power and potential of so-called 'deep learning' in the early 2010s, and which had been built through the hidden work of over 50,000 people who tagged 14 million images (Denton et al., 2021). Data work is also a key ingredient of the development of autonomous vehicles (Schmidt, 2019), voice assistants (Tubaro & Casilli, 2022) and the large language model behind popular chatbot ChatGPT (Perrigo, 2023). Industry sources reckon that demand for data work is growing (Cognilytica, 2022).

Despite their essential role in technology production, these are lower-level tasks characterized by fragmentation and standardization. They constitute peripheral activities that AI producers typically shed to outside suppliers. Digital labour platforms that connect providers to clients, often internationally, are prominent enablers of this 'online outsourcing' (Kuek et al., 2015). A widely known example is Amazon Mechanical Turk, accompanied by several dozen competitors (Kässi et al., 2021) such as Appen, Clickworker, OneForma, Microworkers, and Telus International. With online outsourcing, technology businesses avoid taking direct responsibilities for data workers, while still benefiting from their productive efforts and even imposing quality standards on them. As in other industries (Tomaskovic-Devey & Avent-Holt, 2019), externalised workers operate in very competitive environments that put downward pressure on their remunerations, while their status excludes them from the resources of lead companies.

Nonetheless, the segmented nature of data tasks has raised hopes that even low-skilled people can perform them, with potential to create earning opportunities for disadvantaged populations (Kuek et al., 2015). The possibility to perform these tasks remotely, in an 'online labour market' that crosses national boundaries (Graham & Ferrari, 2022), seems to open this opportunity to emerging and lower-income countries (Datta et al., 2023). The term 'micro-work', often used as a synonym of data work, was indeed coined after 'micro-finance' to highlight the potential of this activity, capable to 'connect poor people to the digital supply chain so that they can earn a living' (Gino & Staats, 2012).

So, do data tasks offer opportunities to more peripheral regions, despite the disruptions they bring to labour markets in core AI-developing countries? Extant evidence is mixed. On the same international platforms, workers are paid differently: average hourly earnings reach US$4.70 in North America and US$3.00 in Europe and Central Asia, but they plummet to US$1.33 in Africa (Berg et al., 2018). These inequalities are partly due to higher worker competition in middle- and lower-income countries, where digital labour supply has grown more strongly (ILO, 2021). They are also partly due to the choices of lead companies to target some of the best-paying tasks to specific groups of workers according to criteria like language and country of residence, with North America being the preferred option.

Exclusion of some countries from many higher-quality tasks drives their workers to do lesser-paid assignments offered by specialized intermediaries in less visible market niches. As an extreme example, Lindquist (2022) and Ong and Cabañes (2019) document the presence in South-East Asia of 'follower factories' and 'click-farms', a type of data work where tasks range from creating fake content to 'liking' and 'sharing' clients' online profiles. Though disruptive of the normal functioning of social media websites, these tasks still contribute to training search and recommendation algorithms. Likewise, captcha-solving tasks serve hackers' dubious goals while also supplying image tags to train computer vision (Morreale et al., 2023). Captcha solvers and click farmers sit in a continuum with 'commercial content moderators' who are hired to filter spam, illicit content, and fabricated data (Roberts, 2019).

The extent to which data tasks offer an opportunity to low-skilled workers is also questionable. The majority of data workers are highly educated, well beyond the requirements of most data tasks (ILO, 2021). Rani and Furrer (2019) see data work as an inefficient use of human capital, especially in lower-income countries where it is scarce. Platforms attract well educated workers who lack suitable local alternatives, but they offer limited opportunities to move up to more complex and challenging tasks, not to mention to improve career prospects.

Growing research also addresses the extent to which data work reproduces, and perhaps expands, informal economy arrangements. Particularly in Latin America, informality is a pillar of the historical division of labour and constitutes a specific form of exploitation and accumulation that supports the peripheral development of global capitalism (Oliveira, 2003). Traditionally, a significant part of the workforce in Argentina, Brazil, and Venezuela, has depended on informal work for subsistence. In this context, platforms may seem to introduce elements of formality through systematic documentation of interactions, measures of performance, and standardisation of communications (Weber et al., 2021). But insofar as platforms deprive workers of rights, and break the guarantees of formal employment, they can be construed as channels that spread informal conditions worldwide. While debates on platformised digital labour in the United States and Europe emphasize threats to fair pay and decent working conditions, 'the "rest of the world" was already living and working informally and precariously' (Surie & Huws, 2023, p. 3).

A rich Latin American literature addresses these ongoing transformations. Beyond any rigid duality between formal and informal work, productive and unproductive labour, Abílio (2020, 2021) describes the progressive shift of risks and costs onto workers as 'informalization'. These workers often move between formal and informal jobs, sometimes holding both simultaneously (Abílio, 2019). Their mixed survival strategies blur the lines between formal and informal work, as well as employment and labour, due to the continuous redefinition of existing occupations. Especially after 2015, the historical process of labour informalization in Latin America has aligned with 'uberization', understood as a new form of control, organization, and management of work (Abílio, 2021). The intensification of labour informalization has absorbed many workers affected by unemployment, low and unequal wages, and progressive erosion of rights. Platforms epitomise this trend, subjecting workers to new forms of control and reducing them to 'on-demand workers, just-in-time workers' (Abílio, 2021, p. 2). In this context, we ask how data work emerges as an alternative source of subsistence, representing new dimensions of informality and/or extending pre-existing forms of informal labor.

## Data work in Latin America

The burgeoning research on digital labour in Latin America has attracted scholarly and policy attention toward the conditions of drivers, delivery couriers and domestic service providers (Haidar et al., 2020), and to a lesser extent toward qualified freelancers like computer programmers and designers (Longo

& Fernández Massi, 2023) but has largely overlooked data work. The large cross-country surveys of Berg et al. (2018) cover data work throughout Latin America, but with relatively few respondents compared to other regions, mostly from Venezuela and secondarily, Brazil. Research by the World Bank (Datta et al., 2023) and the renowned Online Labour Index (Stephany et al., 2021) cover Latin America extensively but conflate data work with other forms of digital labour that can be done remotely, notably qualified freelancing. While these studies suggest the presence of large reservoirs of online workforce in the region, they do not estimate the proportion involved in data work.

A few small-scale, qualitative studies with focus on a single country or case provide additional evidence and help identify key settings and issues. To the best of our knowledge, this limited literature has explored only Argentina, Brazil and Venezuela, suggesting that these three countries constitute an essential starting point for any further analysis. In Argentina, Miceli et al. (2020) and Miceli and Posada (2022) observe data work in a so-called 'impact-sourcing' organization that gives jobs to underprivileged people to annotate images and other digital contents for foreign AI producers. This particular organization offers office space to its workers and manages them directly on behalf of its international clients, thereby resembling a formal body; however, the authors observe that categories and criteria are imposed on workers, who are barely able to develop and exert their subjectivity even when their feedback would improve output quality. More research is thus needed to assess the extent to which workers from a broader range of backgrounds also engage in data work: the large numbers who practice some form of online labour as per Datta et al. (2023) suggest that many of them may be involved in this activity. If so, who are these people, and how do they differ from qualified freelancers on the one end, and from delivery couriers and drivers on the other?

There is more literature on Brazil, where over 50 data work platforms, both national and international (Viana Braz, 2021), offer tasks that range from data preparation for machine learning to more mundane digital support activities (Grohmann & Araújo, 2021; Moreschi et al., 2020), and most prominently click-farming (Grohmann et al., 2022a, 2022b). These studies raise awareness on this phenomenon, suggesting that it may be very widespread and constitute an important, albeit previously overlooked link in the global supply chain of AI and the digital economy. A major claim is that data work is closely intertwined with the historical informality of labour in the country and pursues it through digital tools. Especially click-farming occurs at the intersection of informality and illegality, paying workers to inject various degrees of fake content onto social media. While highly valuable for the insight they bring, these are all small-sized qualitative surveys and case studies that prevent broader comparisons. There is a need for more detail on the people who perform these tasks online, their specificities compared to other countries, and their precise role in AI production.

Venezuela has attracted significant attention in recent years as its workers have flooded international labour platforms after the country's economic and political collapse in the mid-2010s (Johnston, 2022). Some data work platforms for AI amass workforces that are in some cases nearly 75 percent Venezuelan (Schmidt, 2019, 2022). To earn their place in the AI industry, workers have to surmount considerable infrastructural problems like power cuts, slow internet connection, and often outdated devices. They operate in a regulatory vacuum whereby the country' institutions are unable to provide sufficient protections, giving foreign platforms free rein to impose and enforce rules to their advantage (Miceli & Posada, 2022). Left to their own devices, workers depend on informal support from their families and local communities (Posada, 2022). More research is necessary on the identities of these workers and how they maintain these forms of mutual support while also managing competition from their increasingly numerous compatriots.

The present article builds on this literature and harnesses a larger-scale, cross-country, quali-quantitative set of data to generalize extant findings, deepen our knowledge of Latin American data workers, and pinpoint the place of the region in the global production of AI. If both Brazil and Venezuela are large hubs of data work, how do they cope with the geographical, cultural, and especially linguistic distance that separates them from AI producers in the United States and Europe? If the political-economic outlook of Venezuela shapes its participation to global online labour markets, can we expect a similar trend in Argentina, similarly affected by severe economic and financial hardship? If informality plays a pivotal role in the experience of Brazilian data workers, what forms does it take in the other two countries, where it also has a long tradition?

## Methodology

To address these questions, we leverage data from two twin, mixed-method studies of workers who use platforms to do tasks for machine-learning development. The first study targeted data workers from Argentina, Venezuela, and for comparison, all Spanish-speaking Latin American countries. Between December 2020 and January 2021, we launched an online questionnaire, in Spanish, to users of the international platform Microworkers.com, which is US-based but operates worldwide, hosts a large variety of data tasks from AI training to click-farming, and is popular in the region. In the next few months, we conducted follow-up interviews with a sub-sample of questionnaire respondents. Both the questionnaires and interviews were fielded as paid tasks through the platform. In Spring 2022, we conducted a twofold test to ensure the collected data did not idiosyncratically reflect the specific conditions of the COVID-19 pandemic. We launched a follow-up questionnaire to previous Microworkers respondents, while re-fielding the original questionnaire through another international platform, Clickworker. The second Microworkers questionnaire provided evidence of high turnover, though less pronounced in Venezuela and especially in Argentina compared to the rest of Latin America. It appears that COVID-induced restrictions to mobility constituted powerful incentives to platform data work in 2021, while unfavourable macroeconomic conditions maintained these incentives in Argentina and Venezuela, but not elsewhere, in 2022. Finally, to better understand these contextual effects, we conducted more interviews with Argentinean users of Microworkers in Autumn 2023. Overall, we collected 1675 questionnaire responses, of which 220 from Argentina and 214 from Venezuela, plus 286 second-wave responses; and we interviewed 62 workers, 15 of whom from Argentina and 22 from Venezuela.

The second study was conducted in 2022-23, in Portuguese, with Brazilian data workers. We started with an online ethnography of social media groups formed around this new form of labour and grouping workers of different international platforms, and we used it to recruit participants to 15 exploratory interviews. Then, in January 2023, we fielded the same questionnaire that we had used for the rest of Latin America, translated and with some minor adaptations to reflect country specificities, as a task on Microworkers.com. The 477 complete responses that we obtained make it the largest study ever conducted about data work in Brazil.

Because the characteristics and boundaries of platform populations are unknown (Kässi et al., 2021), notions of statistical representativeness hardly apply. Nevertheless, use of the same questionnaire, designed with the same survey software and interface (Limesurvey), and fielded in the same way, ensures comparability across the three countries, with the rest of Latin America as a benchmark. Our relatively large-sized datasets and the inclusion of workers from two different platforms (for the Spanish-language study) provide variation in the data, and triangulation of methods supports interpretation of results. We use questionnaire and interview data jointly, with the former providing a broad overview of the data working universe, while the latter augment the quantitative findings, adding depth and detail based on the experience of data workers, the discourse they use, and their feelings and motivations. We focus on the topics that are closest to the questions addressed in the quantitative analysis, so that each can shed light on the other and put it in context.

The surveys were long, with over 100 questions covering basic socio-demographic information, family situation, education and skills, professional status and experience, income, internet usage, and practices of data work. The qualitative interviews lasted 45 minutes on average and were conducted through a web conferencing system. They were all recorded, transcribed, and accompanied by short debriefing reports by the interviewer(s). After the fieldwork was finished, the interview transcripts and reports were read and excerpts coded. The codes are partly descriptive (types of tasks, personal trajectories) and partly theoretical, inspired by literature (place of informality, access to social protections, the purposes of tasks in AI production, career trajectories and perspectives).

## Analysis

### Argentina

Unlike other Latin American countries, Argentina has protective labour legislation that covers a significant portion of workers, resulting from a strong history of formal wage employment and unionization until the 1970s. At the end of the 20th century, like the global scenario, a process of deterioration of the labour market and the loss of hegemony of formal salaried work became evident, expanding the segment of workers excluded from the protection of labour legislation (Gago, 2011). In 2019, 42.3% of the employed population was engaged in informal and precarious activities (Longo et al., 2023). At the same time, in recent years, in a context of strong macroeconomic instability, the purchasing power of wages fell drastically. Thus, two classic characteristics of the Argentine labour market (formal wage labour and high wages relative to the region) were severely degraded.

The COVID-19 pandemic intensified both processes. Though the unemployment rate commenced a declining trend, reaching a historically low level of 6.3% in the fourth quarter of 2022, surging inflation impoverished the working population (Poy & Alfageme, 2022). Between December 2020 and 2023, real wages exhibited a staggering decline of 19.3%. Inflation is mainly driven by the external constraint with which Argentina has grappled for nearly a decade, characterized by a shortage of dollars to facilitate international payments and progressive depletion of the Central Bank's reserves (Wainer & Belloni, 2022). The exchange control measures that have been implemented to address this challenge, including restrictions on the acquisition of foreign currency at the official exchange rate, have driven the emergence of alternative, parallel exchange rates. One of them stems from the acquisition and disposal of financial securities to access foreign currency and often serves as a benchmark for financial transactions, including those involving virtual wallets. Their trajectories have been similar, with increasing divergence from the official exchange rate. In real terms, i.e., considering the effects of local and external prices, the exchange rate gap began to widen from September 2019, with the financial exchange rate significantly surpassing the official rate. Throughout 2022, the gap amounted to 98% on average, and at its peak in July 2022, the financial exchange rate stood at 136% above the official rate.

In this context, individuals earning income in dollars can capitalize on the opportunity offered by the parallel markets, securing exchange rates considerably more advantageous than the official one. This scenario, coupled with increased familiarity with remote work in the aftermath of the pandemic, has popularized international platforms that remunerate in hard currency, including those for data work. Respondents to our questionnaire indicate that their motivations to engage in this activity are closely linked to the desire to supplement their income, with emphasis on acquiring foreign currency. Table 1 shows that these workers are young (over two thirds are below 35 years of age), mostly male (over two thirds), and highly educated: two out of five have University degrees, well above national averages for the active population, though below the 57% observed by the ILO worldwide (Berg et al., 2018).

**Table 1: Context and data workers' characteristics.**
[a] Source: ILO 2023.
[b] Source: authors' elaboration.

|  |  | **Argentina** | **Brazil** | **Venezuela** |
|---|---|---|---|---|
| **Macro-economic outlook** [a] |  |  |  |  |
| GDP per capita (2022) |  | US$12,932 | US$9,455 | US$3,459 |
| **Data workers' profiles and practices** [b] |  |  |  |  |
| Age | 18-34 | 71% | 71% | 68% |
|  | 35-54 | 27% | 27% | 28% |
|  | 55+ | 2% | 2% | 4% |
| Women |  | 30% | 64% | 36% |
| Education | Secondary or lower | 31% | 44% | 15% |
|  | Short tertiary or post-secondary | 28% | 13% | 22% |
|  | University (Bachelor or higher) | 41% | 43% | 63% |
| Professional status | Salaried employee | 31% | 29% | 19% |
|  | Independent worker | 21% | 21% | 25% |
|  | Student | 27% | 9% | 25% |
|  | Unemployed | 13% | 21% | 14% |
|  | Inactive | 7% | 7% | 11% |
|  | Other | 3% | 13% | 6% |
|  | of which in informal work | NA | 12% | NA |
| Data work as primary income source |  | 31% | 34% | 75% |
| Nb. data tasks per month (average) | Less than 10 | 66% | 81% | 36% |
|  | Between 10 and 20 | 15% | 8% | 13% |
|  | Over 20 | 19% | 11% | 51% |
| Monthly earnings from data work (average) |  | US$83 | US$112 | US$70 |
| Use of earnings from data work | Necessities | 20% | 34% | 68% |
|  | Savings | 55% | 34% | 24% |
|  | Discretionary expenses | 10% | 4% | 2% |
|  | Not earned enough | 15% | 28% | 5% |

This does not make data work a primary source of income, though, in contrast to what happens with physical service platforms: two thirds of Argentineans consider it as supplementary income. Significantly, half of respondents hold salaried or self-employed positions outside these platforms. Within this group, only 7% of salaried workers and 28% of self-employed workers rely on data tasks as their primary source of income. Among those without stable jobs, 1 in 2 unemployed individuals and 1 in 3 students use platforms as their main income source. Thus, Argentinean data workers allocate relatively few hours to this endeavour and, consequently, earn modest incomes. Notably, two thirds of participants engaged in 10 tasks or fewer in the month preceding the survey, and a little less than one fifth performed more than 20 tasks. Average monthly earnings from data work are about US$83, though the distribution is skewed, with values around zero for one third of workers and a few high

earners. Accordingly, the majority of individuals performing data tasks in Argentina utilize their platform-generated income for savings, with only one in five earmarking this money for essential expenses – less that the regional average as per our survey (one third). In Argentina there has been a process of popularisation of financial investments, driven by a number of factors: accelerating inflation, currency gaps, financial digitalisation and a wide diffusion of investment products for 'amateur investors' (Sánchez, 2024). The possibility offered by digital work platforms to generate income in dollars, not declared to local tax authorities, is compatible with investing it in digital financial platforms that are not under local banking regulation.

Why aren't these workers engaging more intensely with data tasks? The financial complexity of getting paid acts as a barrier to entry. To avoid the official exchange rate used in the local banking system, workers have to use virtual wallets, digital dollar accounts, and cryptocurrencies to preserve the value of their dollars as long as possible; when they eventually need to make local expenses that require conversion to Argentine pesos, they must resort to informal channels. The profile of those who use these platforms in Argentina – mostly young men – also matches the profile of those who use digital investment platforms the most, even though the average investment amounts are very low (Sánchez, 2024). More importantly, full-time engagement in platform data work would incur the risk of not getting enough tasks, owing to volatile international demand.

The picture becomes then clearer: data work offers a 'necessary extra income' in dollars to workers whose local wages have been eroded by inflation, but it can rarely be practiced profitably as a full-time activity. Local jobs still constitute the bulk of workers' income, with an average wage of US$ 614 in 2022 for registered salaried employees, and a legal minimum wage – commonly used as a proxy for informal wages – of US$ 174, at the parallel exchange rate. Data tasks are appealing as a complement rather than a substitute of such jobs, and workers benefit most when they are embedded both in the formal economy through their main employment, and in informal markets for money exchange.

*Brazil*

Brazil is an upper middle-income country, the largest economy in Latin America, and the seventh most populous worldwide. Its information technology market is booming, constituting almost half of the whole of Latin America (ABES, 2022). It is one of the largest exporters of outsourced computing services, only behind India, China, and Malaysia, and it hosts some of the highest-ranked universities in information technologies within the region, producing about 50,000 new graduates each year. However, it is also the eight most unequal country in the world (IBGE, 2023), where the richest 5% have as much income as the remaining 95%. Furthermore, Brazil has been undergoing a progressive erosion of labour and social rights in recent years, and it currently has 9.4 million unemployed people, 18% of whom are aged 18-24.

Among the 97.8 million individuals constituting the active workforce, a whopping 39% are engaged in informal labour, a legacy of Brazil's late industrialization in the 20th century. Ineffective regulations, income inequalities, and the unresolved legacy of slavery produce surplus labour in urban areas and low wage share in the population's overall income (Manzano et al., 2023). Economic growth from 1930 to 1980 did not lead to substantial formalization of the labour market, and large portions of the (particularly black, young, and female) population remained excluded. Between 1980 and 2000, the idea that laxer labour laws and more informality could counter unemployment, led to neoliberal reforms, and 57% of newly created salaried jobs were informal. Labor formalization occurred between 2003 and 2014, driven by Workers' Party policies during a period of economic growth. The informality rate decreased from 55% to 41% (Manzano et al., 2023), but after 2015, the boundaries between formal and informal work became less clear. The 2017 Labor Reform introduced the concepts of 'exclusive self-employed worker' and intermittent work, allowing to hire workers as service providers while subjecting them to subordination conditions typical of salaried employment. Including these service providers, most income generation in the country now comes from informal occupations.

While Table 1 shows that the Brazilian data workers in our sample are as young as their Argentinean counterparts (the majority being below 35 years of age), and are equally well educated, with two out of five holding a university degree, they are predominantly women (two thirds). Almost two out of five are unemployed, without professional activity, or in informal work. Career discontinuities are common, in that half of the workers have held at least two informal positions, and over two thirds have held at least three formal jobs throughout their lives. Among those in formal employment, two fifths work part-time, while the global average does not exceed one third (Berg et al., 2018). Their mean monthly household income (US$359), all sources included, stands at about 1.4 times the legal minimum wage, but is almost one third lower than the average income of the general Brazilian population (US$524) and falls short of the estimated monthly living expenses for a single person in cities such as Rio de Janeiro, São Paulo, and Belo Horizonte, where slightly over half of respondents live (US$1038 in 2022).

Zooming in on women is helpful to better understand how data work differs between Brazil and Argentina. Among Brazilians who have no employment outside platform data tasks, almost three quarters are women. Almost two out of five women depend solely on data work platforms for income, compared to little more than one in five men. Slightly over 55% of men are salaried employees, contrasting with 41% of women. Almost two thirds of female respondents are mothers or legal guardians of one or more children, compared to less than half of men. Women spend almost 40% more time on household chores and care duties than men, and they are more numerous to rely on their partners to support their households (62% against 40% respectively).

This evidence suggests that the practice of data work follows different patterns in Argentina and Brazil. The recent inflationary crisis affects each and every resident of Argentina, prompting widespread use of international digital labour platforms as a buffer; but because use of these platforms requires financial and computing skills as well as connections in both the formal and informal economy to achieve profitability, it remains a side activity *de facto* restricted to parts of the country's higher socio-economic strata. The more stable macroeconomic outlook of Brazil makes data tasks unattractive to the better-off, and rather appealing to disadvantaged albeit computer-literate people. Un- and under-employed workers, part-timers, low earners, and mothers without a sufficiently remunerated main job, are more likely to undertake data work to earn a much-needed supplement of income. The uptake of this activity aligns with other forms of digital labour, such as delivery and transportation, though the possibility to do tasks from home makes it more accessible to women with children, representing a continuation of informal work in cities. Overall, data work mirrors the inherited distribution of inequalities in Brazilian society, cumulatively defined by socio-economic status and gender.

Does the practice of data work improve the stakes of this population segment? Evidence is mixed. The average monthly income on platforms is US$112, for an effort of about 15 hours and half of work a week, though the distribution of both variables is again markedly skewed. Slightly over one third of respondents depend on data work as their main source of income, though this figure increases to almost 40% for women, and diminishes to 24% for men. Interestingly, women earn slightly more than men, partly because they enter platforms more frequently. They often do so during business hours for repeated short intervals, while men usually connect for longer durations after their main job.

Despite the financial gain they get from this activity, workers complain, and interviews were a helpful way of capturing their perspective. Beyond known grievances, for example about the repetitive nature of tasks, they raise issues that escaped previous studies set in the United States and Europe. To begin with, task replenishment occurs in the time zones of the clients' countries, putting Brazilians at a disadvantage when working on international projects: unsurprisingly, more than one fourth of respondents start working in the early morning hours. Additionally, minimal interactions with same-language peers worsens the effects of fatigue and tiredness. While several discussion groups dedicated to this topic exist on WhatsApp, Telegram, and Facebook, they are informal and often not authorized by platforms. Only 22% of workers engage in them, and the majority use them passively.

*Venezuela*

A major oil exporter since the 1970s, Venezuela achieved ambitious social goals under the presidency of Hugo Chávez in 1999-2013, among others reducing child mortality, extending life expectancy, expanding education access, and spreading digital access and literacy. But fluctuations in oil prices since the mid-2010s, combined with political instability and sanctions, have plunged the country into a deep economic crisis. High emigration, declining birth rates, and increasing mortality (ENCOVI, 2022) have shrunk the population, while recent improvements are modest, encompassing a drop in the annual inflation rate from 600% in 2021 to 305.7% in 2022, and a reduction in poverty from 90.9% to 81.5%. Its inflation rate is still the highest in the world, inequality is high, the number of young people (3-24 years old) in education diminishes, and the average wage of US$138 is insufficient to buy the basic household food basket, penalizing especially public-sector employees and retirees (OVF, 2023).

In these conditions, data work through international platforms appears as a channel to access hard currency, and even the smallest tasks compare favourably to local wages paid in domestic money. The massive inflow of Venezuelans on platforms, documented in the above-cited literature, is a response to these challenging circumstances. Interviews with workers from different countries hint that Venezuelans have a reputation for accepting even the lowest-paying tasks – like captcha-solving at rates of US$0.50-1.50 per 1000 captchas, on (usually Russia-based) specialized platforms. Their age profile is similar to the cases of Argentina and Brazil, with almost 70% under 35 years of age (Table 1). As in Argentina, the majority (almost two thirds) are men. Their educational profile is more extreme, with almost two thirds holding a university degree, thus exceeding the world average of 57% (Berg et al., 2018); their degrees are often in technical disciplines like engineering and computing. Compared to the other two countries, fewer people (less than 20%) are in salaried employment beyond data tasks. The most striking difference, however, is that platform data work is the main source of income for three quarters of Venezuelan data workers, and earnings are overwhelmingly used to purchase necessities. Data work is not a mere source of supplementary income in Venezuela, but the essential core of people's activity and their primary hope of survival. The vanishing purchasing power of the domestic currency makes it preferable to undertake data tasks full time rather than combining them with an off-platform salaried job – unlike the above-discussed case of Argentina. This explains that over half of study participants report doing more than 20 tasks a month, much above their Argentinean and Brazilian counterparts, and that many start very early in the morning to synchronize with clients' business hours. While their average monthly earnings are somewhat lower than those observed in the other two countries, partly owing to willingness to do even very low-paid tasks, Venezuelans are under-represented within the group with no earnings.

These data suggest that some sort of virtual brain drain is occurring in Venezuela, where an elite workforce of university-educated engineers and scientists, increasingly scarce in a country whose education system is deteriorating, is devoting time and energy to low-priced data tasks on international platforms rather than engaging in the local economy. Their skills are rarely used to perform these tasks, which are repetitive and often unchallenging, but to surmount the numerous obstacles that could bar successful participation. Interviews clearly indicate that these workers do research on clients to better understand their expectations and respond accordingly, as a way to counter the adverse effects of their aging computing equipment, slow internet connections, and frequent power outages. Workers invest in cryptocurrencies and liaise with intermediaries in the informal market to convert their online earnings into local currency, so as to mitigate the negative effects of an unfavourable official exchange rate and of financial restrictions triggered by international sanctions. They write bits of code, programme hotkeys, and use automated tools (such as translators) extensively to do tasks more quickly and therefore beat the competition – tighter in Venezuela than elsewhere due to the large numbers. They sometimes take the risk of using Virtual Private Networks (VPNs), banned by most platforms, to pose as residents of neighbouring countries and access a broader, and allegedly better paid, range of tasks. Workers with the competencies and willingness-to-learn to address these different challenges stay afloat in this market, while traditionally disadvantaged workers with less available time will fall backward – typically, women with care duties.

Nevertheless, competition coexists with cooperation, and Venezuelans are more numerous than any other nationality in our sample to interact with other people who also do data tasks, and to join online peer groups on Telegram, WhatsApp, and Facebook. They are also among the most active initiators of such groups and have become a reference for other Spanish speakers. Online groups are important vehicles to build and share knowledge that helps members navigate international platforms and their challenges. Through these informal interactions, data work has become a nationally-widespread activity that contributes to shaping people's everyday life – unlike anywhere else in Latin America.

## Discussion and conclusions

The preceding cross-country analysis unveils the central place of Latin America in the globalised AI industry and raises the question of the policies that may ensure visibility, recognition, and fair remunerations to this important yet often disregarded contribution. If the emerging field of AI ethics mostly focuses on the effects of the *deployment* of these technologies in society, the study of data work acts as a forceful reminder that ethical reflection should extend to the underpinning *production* processes. Shaped by the outsourcing and offshoring practices that have become prominent in recent decades, AI production is a political-economic phenomenon intertwined with globalization.

We have seen that data work occurs at the margins of AI production, where costs can be maximally compressed. In crisis-stricken countries like Venezuela and to a lesser extent Argentina, participation to this activity is driven by demand for hard currency and therefore serves foreign platforms and clients. This dynamic aligns with Couldry and Mejias's (2019) concept of 'data colonialism,' which frames the extraction of data and labour from the Global South as a continuation of historical colonial processes, now manifesting through digital platforms and AI technologies. However, while most flows of data work reproduce legacy colonial patterns connecting Latin America to the United States and Europe, others (Venezuela to Russia) mirror the emergence of new, and potentially conflicting, centres of power. The Latin American case shows how global AI production thrives on countries' economic hardship and historical inequalities, while also being shaped by political tensions.

If procurement of data for AI has gone global, the earth is not flat. Data workers in Argentina, Brazil, and Venezuela do not experience the same conditions as their counterparts in richer countries. The difficulties they experience to convert their online earnings into local currency distinguish them sharply from US workers who can simply wire the money to their bank account. To serve overseas AI producers, Latin American data workers must synchronize with foreign business hours, work in a second language (English), learn to efficiently use automated translators and other tools not needed elsewhere.

AI data work redefines digital and social inequalities within Latin America too. In Argentina and Venezuela, tough competition and the need for both computing and financial skills favour the emergence of 'elite' workers who are overwhelmingly young, male, and highly educated. They earn their place in this planetary market but their large numbers keep remunerations down. In Argentina, they are insufficient to earn a living, and in Venezuela, the ongoing virtual brain drain does not open any career prospects. In more stable but very unequal Brazil, data work is more often taken up by relatively underprivileged segments of the population (though sufficiently equipped and computer-literate), including informal, unemployed, and underemployed workers as well as women with care duties. Nonetheless, earnings are not enough to raise them to higher socio-economic positions.

In all three countries, data work for AI is interwoven with the historically prevalent informal economy. In Argentina, participation in informal exchange rates and black markets is essential to benefit from data work. In Brazil, the informal sector provides abundant and willing workforce to platforms. In Venezuela, informal interactions spread knowledge about digital labour and platforms, enabling broader participation. These three cases align with the literature about the blending of formal and informal economies, with workers frequently shifting between the two. They confirm the already-noted links between platformisation and informalisation, while also revealing that these connections extend beyond platform-based data work narrowly interpreted. There are spillovers into other sectors, like finance, with variations depending on country and context. These cross-sector dependencies tie informality to inequality: though informal work is socially devalued, for example, access to informal exchange houses is often easier for relatively privileged groups. This adds a layer of complexity to the effects of digitalisation on informalisation.

Our study has policy implications at global and local levels. Globally, it calls for more attention to the conditions of production of AI, especially workers' rights and pay. One instrument to achieve this are due diligence legislations which, in European countries at least, require large companies to systematically audit their providers and contractors, wherever located, for respect of labour legislation and human rights. At the time of writing this article, application to the technology sector is still hesitant, and significant effort is needed to define standards and guidelines that help businesses adopt this regulation more smoothly and systematically. At local level, it is important to devise solutions for the recognition of skills and experience of data workers, in ways that may support their further professional development and trajectories. Support of initiatives for socialization and knowledge sharing, like online groups, would also contribute positively to quality of life at work, give more visibility to workers in public debates, and possibly facilitate some initial forms of worker organization.

# References

ABES (Brazilian Software Association). (2022). *Brazilian Software Market: Scenario and Trends*, (Report). Associação Brasileira das Empresas de Software.

Abílio, L.C. (2019). Uberização: Do empreendedorismo para o autogerenciamento subordinado. *Psicoperspectivas. Individuo y Sociedad*. 18(3):1-11.

Abílio, L.C. (2020). Plataformas digitais e uberização: Globalização de um Sul administrado? *Contracampo: Brazilian Journal of Communication*. 39(1): 12-26.

Abílio, L.C. (2021). Uberization: informalization and the just-in-time worker. *Trabalho, Educação e Saúde*, 19(1): e00314146. https://doi.org/10.1590/1981-7746-sol00314

Berg, J., Furrer, M., Harmon, E., Rani, U., & Silberman, M.S. (2018). *Digital Labor Platforms and the Future of Work: Towards Decent Work in the Online World*, (Report). ILO.

Cognilytica. (2022). *Data Labeling Market: Research Snapshot Dec. 2021*, (Report). Cognilytica.

Couldry, N., & Mejias, U.A. (2019). *The Costs of Connection: How Data is Colonising Human Life and Appropriating It for Capitalism*. Stanford University Press.

Datta, N., & Chen, R. (with Singh, S., Stinshoff, C., Iacob, N., Simachew Nigatu, N., Nxumalo, M., Klimaviciute, L., et al.) (2023). *Working Without Borders: The Promise and Peril of Online Gig Work,* (Report). World Bank.

Denton, E., Hanna, A., Amironesei, R., Smart, A., & Nicole, H. (2021). On the genealogy of machine learning datasets: A critical history of ImageNet. *Big Data & Society*, 8(2), https://doi.org/10.1177/20539517211035955

Ekbia, H.R., & Nardi, B.A. (2017). *Heteromation, and Other Stories of Computing and Capitalism*. MIT Press.

ENCOVI (Encuesta Nacional de Condiciones de Vida). (2022). Universidad Católica Andrés Bello. URL: https://www.proyectoencovi.com/

Gago, V. (2011). De la invisibilidad del subalterno a la hipervisibilidad de los excluidos. Un desafío a la ciudad neoliberal. *Nómadas*, 35, 49-63.

Gino, F., & Staats, B. (2012). The microwork solution. *Harvard Business Review*, 90(12), 92-96.

Graham, M., & Ferrari, F. (eds.) (2022). *Digital Work in the Planetary Market*. MIT Press.

Grohmann, R., & Araújo, W.F. (2021). Beyond Mechanical Turk: The work of Brazilians on global AI platforms. In P. Verdegem (Ed.), *AI for Everyone? Critical Perspectives* (pp. 247-266). University of Westminster Press.

Grohmann, R., Aquino, M.C., Rodrigues, A., Matos, É., Govari, C., & Amaral, A. (2022a). Click farm platforms: An updating of informal work in Brazil and Colombia. *Work Organisation, Labour & Globalisation*, 16(2), 7-20, https://doi.org/10.13169/workorgalaboglob.16.2.0007.

Grohmann, R., Pereira, G., Guerra, A., Abílio, L.C., Moreschi, B., & Jurno, A. (2022b). Platform scams: Brazilian workers' experiences of dishonest and uncertain algorithmic management. *New Media & Society*, 24(7), 1-30, https://doi.org/10.1177/14614448221099.

Haidar, J., Menéndez, N.D., & Bordarampé, G. (2021). *Las plataformas de reparto en Argentina: Entre el cambio de gobierno y la pandemia*, (Método CITRA vol. 8). CONICET/UMET.

IBGE (Instituto Brasileiro de Geografia e Estatística). (2023). *PNAD Contínua - Pesquisa Nacional por Amostra de Domicílios Contínua*. Segundo trimestre de 2023.

ILO (International Labor Organization). (2021). *World Employment and Social Outlook 2021: The Role of Digital Labor Platforms in Transforming the World of Work*, (Report). ILO.

ILO (International Labor Organisation). (2023). *Employment Situation in Latin America and the Caribbean: Towards the Creation of Better Jobs in the Post-pandemic Era*, (Report). ILO.

Johnston, H. (2022). In search of stability at a time of upheaval: Digital freelancing in Venezuela. In M. Graham & F. Ferrari (Eds.), *Digital Work in the Planetary Market* (pp. 157-173). MIT Press.

Kässi, O., Lehdonvirta, V., & Stephany, F. (2021). How many online workers are there in the world? A data-driven assessment [version 4; peer review: 4 approved]. *Open Research Europe*, *153*, https://doi.org/10.12688/openreseurope.13639.4.

Kuek, S.C., Paradi-Guilford, C.M., Fayomi, T. Imaizumi, S., & Ipeirotis, P. (2015). *The Global Opportunity in Online Outsourcing*, (Report). World Bank.

Lindquist, J. (2022). 'Follower factories' in Indonesia and beyond: automation and labor in a transnational market. In M. Graham & F. Ferrari (Eds.) *Digital Work in the Planetary Market* (pp. 59-75). MIT Press.

Longo, J., Busso, M., & Fernández Massi, M. (2023). Trabajar en Plataformas en Argentina: Usos y valoraciones de esta nueva modalidad de trabajo. *Trabajo y Sociedad*, 24(41), 281-297.

Longo, J., & Fernández Massi, M. (2023). Plataformas de servicios virtuales: Un análisis de los perfiles de quienes trabajan de forma remota desde la Argentina. *Papeles de Trabajo*, 17(32), 99-122.

Manzano, M., Krein, J.D., Abílio, L. (2023). A dinâmica da informalidade laboral no Brasil nas primeiras duas décadas do século XXI. In: R. Antunes (Ed.) *Icebergs à deriva: o trabalho nas plataformas digitais* (pp. 211-229). Boitempo.

Miceli, M., Schuessler, M., & Yang, T. (2020). Between subjectivity and imposition: Power dynamics in data annotation for computer vision. In *Proceedings of the ACM on Human-Computer Interaction*, 4, CSCW2, 115, https://doi.org/10.1145/3415186.

Miceli, M., & Posada, J. (2022). The data-production dispositif. In *Proceedings of the ACM on Human-Computer Interaction*, 6, CSCW2, https://doi.org/10.1145/3555561.

Milberg, W., & Winkler, D. (2013). *Outsourcing Economics: Global Value Chains in Capitalist Development*. Cambridge University Press.

Moreschi, B., Pereira, G., & Cozman, F.G. (2020). The Brazilian workers in Amazon Mechanical Turk: dreams and realities of ghost workers. *Contracampo: Brazilian Journal of Communication*, *39*(1), 44-64.

Morreale, F., Bahmanteymouri, E., Burmester, B., Chen, A., & Thorp, M. (2023). The unwitting labourer: extracting humanness in AI training. *AI & Society*, https://doi.org/10.1007/s00146-023-01692-3.

Oliveira, F. de. (2003). *Crítica à razão dualista/ O ornitorrinco.* Boitempo.

Ong, J.C., & Cabañes, J.V. (2019). When disinformation studies meets production studies: Social identities and moral justifications in the political trolling industry. *International Journal of Communication*, *13*, https://ijoc.org/index.php/ijoc/article/view/11417.

OVF (Observatorio Venezolano de Finanzas). URL https://observatoriodefinanzas.com/

Perrigo, B. (2023). OpenAI used Kenyan workers on less than $2 per hour to make ChatGPT less toxic. *Time*, 18 Jan., https://time.com/6247678/openai-chatgpt-kenya-workers/.

Posada, J. (2022). Embedded reproduction in platform data work. *Information, Communication & Society*, *25*(6), 816-834, https://doi.org/10.1080/1369118X.2022.2049849.

Poy, S., & Alfageme, C. (2022). Trabajadores pobres en tiempos de pandemia (2019-2021). In A. Salvia, S. Poy, & J. Plá (Eds.), *La sociedad argentina en la pospandemia: Radiografía del impacto del covid-19 sobre la estructura social y el mercado de trabajo urbano* (pp. 125–140). Siglo XXI.

Rani, U., & Furrer, M. (2019). On-demand digital economy: Can experience ensure work and income security for microtask workers? *Jahrbücher für Nationalökonomie und Statistik*, *239*(3), 565-597, https://doi.org/10.1515/jbnst-2018-0019.

Roberts, S.T. (2019). *Behind the Screen: Content Moderation in the Shadows of Social Media*. Yale University Press.

Ross, J., Irani, L., Silberman, M.S., Zaldivar, A., & Tomlinson, B. (2010). Who are the crowdworkers? Shifting demographics in Mechanical Turk. In *CHI '10 Extended Abstracts on Human Factors in Computing Systems*, 2863-2872, https://doi.org/10.1145/1753846.1753873.

Sánchez, M.S. (2024). Cuando las inversiones se popularizan. Finanzas digitales e inversores amateurs en Argentina. *Estudios sociológicos*, *42*, 1-26, https://doi.org/10.24201/es.2024v42.e2531.

Schmidt, F.A. (2019). *Crowdproduktion von Trainingsdaten: Zur Rolle von Online-Arbeit beim Trainieren autonomer Fahrzeuge*, (Report). Hans-Böckler-Stiftung.

Schmidt, F.A. (2022). The planetary stacking order of multilayered crowd-AI systems. In M. Graham & F. Ferrari (eds.) *Digital Work in the Planetary Market* (pp. 137-155). MIT Press.

Stephany, F., Kässi, O., Rani, U., & Lehdonvirta, V. (2021). Online Labour Index 2020: New ways to measure the world's remote freelancing market. *Big Data & Society*, *8*(2), https://doi.org/10.1177/20539517211043240.

Surie, A., & Huws, U. (2023). Platformization and informality: Pathways of change, alteration, and transformation. In A. Surie & U. Huws (Eds.) *Platformization and Informality: Dynamics of Virtual Work*. Palgrave Macmillan, https://doi.org/10.1007/978-3-031-11462-5_1.

Tomaskovic-Devey, D., & Avent-Holt, D. (2019). *Relational Inequalities: An Organizational Approach*. Oxford University Press.

Tubaro, P., Casilli, A.A., & Coville, M. (2020). The trainer, the verifier, the imitator: Three ways in which human platform workers support artificial intelligence. Big Data & Society, 7(1), https://doi.org/10.1177/2053951720919776.

Tubaro, P., & Casilli, A.A. (2022). Human listeners and virtual assistants: privacy and labor arbitrage in the production of smart technologies. In M. Graham & F. Ferrari (Eds.) Digital Work in the Planetary Market (pp. 175-190). MIT Press.

Viana Braz, M. (2021). Heteromação e microtrabalho no Brasil. Sociologias. 23(57), 134-172, https://doi.org/10.1590/15174522-111017.

Wainer, A., & Belloni, P. (2022). Balance-of-Payments constraints as the key to dependency: The case of Argentina. *Latin American Perspectives*, *49*(2), 144-162, https://doi.org/10.1177/0094582X211069556.

Weber, C.E., Okraku, M., Mair, J., & Maurer, I. (2021). Steering the transition from informal to formal service provision: labor platforms in emerging-market countries. *Socio-Economic Review*, *19*(4), 1315-1344, https://doi.org/10.1093/ser/mwab008.

Wood, A.J., Graham, M., Lehdonvirta, V., & Hjorth, I. (2019). Networked but commodified: The (dis)embeddedness of digital labor in the gig economy. *Sociology*, *53*(5), 931-950, https://doi.org/10.1177/0038038519828906.